
%
%
%
%
\input harvmac.tex


\lref\ka {For a review see I.R. Klebanov and A. Pasquinucci, lectures given at
Trieste Summer School of Theoretical Physics, 1992, hepth/9210105.}
\lref\rev {For a review see I.R. Klebanov, in {\it String Theory and
Quantum Gravity '91} World Scientific 1992,
and references therein.}
\lref\lee {J.C. Lee Phys. Lett. {\bf B337} (1994) 69;
Phys. Lett. {\bf B326} (1994) 79;
Prog. Th. Phys. Vol. 91 No.{\bf 2} (1994) 353.}
\lref\post{A.M. Polyakov, Mod. Phys. Lett. {\bf A6} (1991) 635.}
\lref\kp {I.R. Klebanov and A.M. Polyakov, Mod. Phys. Lett. {\bf A6} (1991)
3273.}
\lref\lz{B. Lian and G. Zuckman, Phys. Lett {\bf B266} (1991) 21.}
\lref\wit{E. Witten, Nucl. Phys. {\bf B373} (1992) 187;
E. Witten and B. Zweibach, Nucl. Phys. {\bf B377} (1992) 55.}
\lref\aj{J. Avan and A. Jevicki, Phys. Lett. {\bf B266} (1991) 35,
{\bf B272} (1991) 17.}
\lref\poly{A.M. Polyakov, Princeton preprint PUPT-1289,
Lectures given at 1991 Jerusalem Winter School,
{\it Jerusalem Gravity 1990}, 175.}
\lref\gkn{D.J. Gross, I.R. Klebanov and M.J. Newman, Nucl. Phys.
{\bf B350} (1991) 621;
D.J. Gross and I.R. Klebanov, Nucl. Phys. {\bf B352} (1991) 671;
K. Demetrifi, A. Jevicki and J.P. Rodrigues, Nucl. Phys. {\bf B365} (1991)
499 ;
U.H. Danielson and D.J. Gross, Nucl. Phys. {\bf B366} (1991) 3. }
\lref\gold{Goldstone, unpublished; V.G. Kac, in {\it Group Theoretical
Methods in Physics, Vol 94}, Springer-Verlag, (1979);
G. Segal, Comm. Math. Phys. {\bf 80} (1981) 301.}
\lref\eky{T. Eguchi, H. Kanno and S. Yang, Phys. Lett {\bf B298}
(1993) 73.}

\font\blackboard=msbm10 \font\blackboards=msbm7
\font\blackboardss=msbm5
\newfam\black
\textfont\black=\blackboard
\scriptfont\black=\blackboards
\scriptscriptfont\black=\blackboardss
\def\blackb#1{{\fam\black\relax#1}}


\def\del {{\partial}}
\def\half {{1 \over 2 }}
\def\ket {{\rangle}}
\def\ZZ {{ \blackb Z }}


\Title{\vbox{\baselineskip12pt\hbox{NHCU-HEP-94-33}
    \hbox{November 1994}
    \hbox{hep-th/9412095}
    }}
{\vbox{\centerline{Discrete Gauge States and}
	\vskip 2mm\centerline{$w_\infty$ Charges in c=1 2d Gravity}}}

\vskip .5cm
\centerline{Tze-Dan Chung\footnote{$^\dagger$}
{\it email: chung@math.nctu.edu.tw} and Jen-Chi Lee\footnote{$^*$}
{\it email: jclee@twnctu01.bitnet}}
\bigskip\centerline{\it Department of Electrophysics}
\centerline{\it and Institute of Physics}
\centerline{\it National Chiao Tung University}
\centerline{\it Hsinchu, Taiwan 30050, R.O.C.}

\vskip 2cm
\rm
\noindent

\centerline{\bf Abstract}

We give a general formula for gauge states at the discrete momenta
in Liouville theory. These discrete gauge states carry the $w_\infty$ charges.
As in the case of the 26D (or 10D)
string theory, they are decoupled from the correlation functions and
can be considered as the symmetry parameters in the old covariant
quantization of the theory.

\Date {}


\newsec {Introduction}

For the past few years, toy 2D string model (or c=1 2d gravity) \refs{\rev}
has been an important
laboratory to study non-perturbative information of string theory.
In the continuum Liouville approach \refs{\ka},
in addition to the massless tachyon mode,
an infinite number of massive discrete momentum physical degrees of freedom
were discovered \refs{\gkn}\refs{\post}
and the target space-time $w_\infty$ symmetry
and Ward identities were then identified \refs{\aj}\refs{\kp}\refs{\wit}.
This important high energy ($\alpha' \to \infty$)
structure turns out to be impossible to be extracted in the 26D (or 10D)
string theory due to the high dimensionality of the
space-time. However, some interesting progresses for these high dimensional
string theories have been made by using two types of gauge states (physical
zero norm states) in the spectrum \refs{\lee}.
 For example, the massive inter-particle
broken symmetries and Ward identities for the first few levels
were demonstrated although the
symmetry algebra is still difficult to identify.
In contrast to the BRST quantization used in the 2D case
\refs{\wit}\refs{\lz},
these were done in the old covariant quantization of the theory.

In this letter, we will derive the $w_\infty$ structure from the gauge states
point of view in the old covariant quantization scheme. This is in parallel
with the works of \refs{\kp} and \refs{\wit} where the ground ring structure
of ghost number zero operators were identified in the BRST quantization.
Moreover, the results we obtained will justify the idea of gauge states used
in the 26D (or 10D) theories as discussed above. Unlike the discrete Polyakov
states, we will find that there is still an infinite number of continuum
momentum gauge states in the massive levels of the 2D spectrum and it is very
difficult to give a general formula for them just as in the case of 26D theory
\refs{\lee}.
However, as far as the dynamics of the theory is concerned, only those gauge
states with Polyakov's discrete momenta are relevant. This is because all
other gauge states are trivially decoupled from the correlation functions
due to kinematic reason. Hence, we will only identify all discrete gauge
states (DGS) in the spectrum. The higher the momentum is, the more numerous
the DGS are found. In particular, we will give an explicit formula for one
such set of DGS in terms of Schur polynomials. Finally, we can show that
these DGS carry the $w_\infty$ charges and serve as the symmetry parameters of
the theory. This is in complete analogy with, e.g., Gupta-Bleuler quantization
of QED where gauge state $\theta (x)$
serve as the $U(1)$ parameter of the theory.

\newsec{Gauge States in 2D Gravity}

We consider the two dimensional critical string action \refs{\ka}
\eqn\act{
S={1\over 8 \pi} \int d^2\sigma {\sqrt{\hat g}} [ g^{\mu\nu}
(\del_\mu X \del_\nu X + \del_\mu \phi \del_\nu \phi)
-Q {\hat R} \phi ] }
with $\phi$ being the Liouville field.
For $c=1$ theory $Q$, which represents the background charge
of the Liouville field, is set to be $2\sqrt2$ so that the total anomalies
cancels that from ghost contribution.

For simplicity here we consider only one of the chiral sectors, while
the other sector (denoted by $\bar z$) is the same.
The stress energy tensor is
\eqn\seten{T_{zz} = - \half (\del_z X)^2 - \half (\del_z \phi)^2
- \half Q \del_z^2 \phi .}

If we define the mode expansion of $X^\mu = (\phi,X)$ by
\eqn\xpmod{
\del_z  X^\mu = - \sum_{n=-\infty}^{\infty}z^{-n-1}
(\alpha_n^0 , i \alpha_n^1 ),}
with the Minkowski metric $\eta_{\mu\nu}= \pmatrix{-1& 0 \cr 0 &1 \cr}$,
$Q^\mu=(2{\sqrt 2} , 0)$ and the zero mode
$\alpha^{\mu}_0 = f^\mu = (\epsilon, p )$, we find the Virasoro
generators
\eqn\vir{\eqalign{
L_n =& ( {n+1 \over 2} Q^\mu + f^\mu) \alpha_{\mu,n}
+ \half \sum_{k \neq 0} : \alpha_{\mu,-k} \alpha_{n+k}^\mu :
\qquad n \neq 0 \cr
L_0 =& \half (Q^\mu + f^\mu) f_\mu + \sum_{k=1}^{\infty}
: \alpha_{\mu,-k} \alpha_k^\mu : .}}

The vacuum $|0 \ket$ is annihilated by all $\alpha_n^\mu$ with $n>0$.
In the old covariant quantization,
physical states $|\psi \ket $ are those satisfy the condition
\eqn\phyc{\eqalign{
L_n |\psi \ket &=0 \qquad for \qquad n>0 \cr
L_0 |\psi\ket  &= |\psi\ket .}}
One can easily check that the two branches of massless ``tachyon''
\eqn\tahcbr{T^{\pm}(p) = e^{ipX+(\pm |p| -{\sqrt2}) \phi} }
are positive norm physical states. In the ``material gauge'' \refs{\kp},
it was also known that there exist discrete states \refs{\gkn} \refs{\gold}
($J=\{0,\half,1...\}$ and $M=\{ -J,-J+1,...J\}$)
\eqn\disc{\psi_{J,M}^{(\pm)} \sim (H_-)^{J - M} \psi_{J,J}^{(\pm)}
\sim (H_+)^{J + M} \psi_{J, -J}^{(\pm)} ,}
which are also positive norm physical states.
In \disc\  $H_\pm = \int {dz\over 2 \pi i} T^+({ \pm \sqrt2})$ are the
zero modes of the ladder operators of the $SU(2)$ Kac-Moody
currents at the self-dual radius in $c=1$ 2d conformal field theory and
$\psi_{J,\pm J}^{(\pm)}= T^{(\pm)}({{ \pm \sqrt 2}J})$.
These exhaust all positive norm
physical states. In this letter we are interested in the
discrete gauge states (DGS), i.e., the zero norm physical states
at the same discrete momenta as those states in \disc .
We thus no longer restrict ourselves in the ``material'' gauge,
and the Liouville field $\phi$ will play an important role in the
following discussions.

In general, there are two types of gauge states,

Type I:
\eqn\gauone{
|\psi \ket = L_{-1} |\chi \ket \qquad where \qquad L_m |\chi \ket =0
\qquad m \geq 0 }

Type II:
\eqn\gautwo{\eqalign{
|\psi \ket =\left(L_{-2} + {3\over2}L_{-1}^2 \right)|\tilde\chi \ket
\qquad where \qquad &L_m |\tilde\chi \ket =0
\qquad m > 0 \cr
& (L_0+1) |\tilde\chi \ket = 0 }}
They satisfy the physical state conditions \phyc , and have zero norm.
It is important to note that \gautwo\ is a gauge state only when
$Q=\sqrt{25-c \over 3}$, while the states in \gauone\ are insensitive
to this condition.
In this section we will explicitly calculate the gauge states at the
two lowest mass levels.
At mass level one (i.e. spin one), $ f_\mu (f^\mu +Q^\mu) = 0$,
only gauge states of type I are found:
$f_\mu \alpha^\mu_{-1} |f \ket$,
where $ |f \ket = :e^{ip X + \epsilon \phi} :|0 \ket$.
The DGS $G^-_{1,0} = :\del\phi e^{-2\sqrt2 \phi}:|0 \ket $
corresponds to the momentum of $\psi^-_{1,0}$. There is no corresponding
DGS for $\psi^+_{1,0}$.

At mass level two, $ f_\mu (f^\mu +Q^\mu) = -2$, if
$e_\mu (f^\mu + Q^\mu) =0$ then the type I gauge state is
\eqn\ogauo{
|\psi\ket = [\half(f_\mu e_\nu + e_\mu f_\nu )\alpha_{-1}^\mu \alpha_{-1}^\nu
+ e_\mu \alpha_{-2}^\mu ]|f\ket ,}
while the type II state is
\eqn\ogaut{|\psi\ket = \half [( 3 f_\mu f_\nu +  \eta_{\mu\nu} )
\alpha_{-1}^\mu \alpha_{-1}^\nu +
(5f_\mu - Q_\mu )\alpha_{-2}^\mu] |f\ket .}

The DGS corresponding to $\psi^-_{{3\over2},\pm\half}$
are $G^-_{{3\over2},\pm\half}$ :

(type I)
\eqn\dgtwo{ G^{-(1)}_{{3\over2},\pm\half} \sim \left[
\pmatrix{ {5\over2}& \pm {3\over2} \cr
	   \pm {3\over2} & \half \cr }
\alpha_{-1}^\mu \alpha_{-1}^\nu
+ \pmatrix { {1 \over \sqrt 2} \cr
	\pm  {1 \over \sqrt 2} \cr} \alpha_{-2}^\mu \right]
|f_\mu =(-{5\over2},\pm\half) \ket }

(type II)
\eqn\dgtwot{ G^{-(2)}_{{3\over2},\pm\half} \sim \half \left[
\pmatrix{ {73\over2}& \pm {15\over2} \cr
	   \pm {15\over2} & {5\over2} \cr }
\alpha_{-1}^\mu \alpha_{-1}^\nu
+ \pmatrix { {29 \over \sqrt 2} \cr
	\pm  {5 \over \sqrt 2} \cr} \alpha_{-2}^\mu \right]
|f_\mu =(-{5\over2},\pm\half) \ket .}
Note that a linear combination of these two states produces a
``pure $\phi$ DGS'':
\eqn\pgstwo{G^-_{{3\over2},\pm\half} \sim
[(\del \phi)^2 - {1 \over \sqrt 2} \del^2 \phi]
e^{\pm {i\over \sqrt 2}X - {5\over \sqrt2} \phi} |0\ket .}

The gauge states corresponding to discrete momenta of
$\psi^+_{{3\over2},\pm\half}$ are degenerate, i.e.,
the type I and type II gauge states are linearly dependent:
\eqn\dgtwopl{G^+_{{3\over2},\pm\half} \sim \left[
\pmatrix{ -\half& \pm {3\over2} \cr
	   \pm {3\over2} & -{5\over2} \cr }
\alpha_{-1}^\mu \alpha_{-1}^\nu
+ \pmatrix { {1 \over \sqrt 2} \cr
	\mp  {5 \over \sqrt 2} \cr} \alpha_{-2}^\mu \right]
|f_\mu =(\half, \pm\half) \ket .}
There is no ``pure $\phi$ DGS'' here. In general, the $\psi^+$ sector
has fewer DGS than the $\psi^-$ sector at the same discrete momenta,
as a result, the ``pure $\phi$ DGS'' only arise
at the minus sector. This fact is related to the degeneracy of the DGS in
the plus sector. Historically
the $\psi^+$ sector discrete states arise when one considers the ``singular
gauge'' transformation constructed from the difference of the
two plus gauge states \refs{\gkn} \refs{\poly}.

\newsec{Generating the Discrete Gauge States}

In this section, we will give a general formula for the DGS. In
general, there are many DGS for each discrete momentum. The higher the
momentum is, the more numerous the DGS are found.
We first express the discrete states in \disc\ in terms of Schur
polynomials, which are defined as follows:
\eqn\schdef{
Exp \left ({\sum_{k=1}^{\infty} a_k x^k}\right)
= \sum_{k=0}^{\infty} S_k \left( \{a_k\} \right) x^k}
where $S_k$ is the Schur polynomial, a function of $\{a_k\} = \{a_i :
i \in \ZZ_k \} $.
Performing the operator products in \disc\ ,
the discrete states $\psi_{J,M}^\pm$ can be written as
\eqn\psieva {\eqalign{
\psi_{J,M}^\pm \sim & \prod_{i=1}^{J-M} \int {dz_i \over 2\pi i} z_i^{-2J}
\prod_{j<k}^{J-M} (z_j-z_k)^2 \cr
& Exp \left[ \sum_{i=1}^{J-M}\big[ -i \sqrt 2 X(z_i)\big] +
\sqrt 2 \big(iJX(0) +
(-1 \pm J) \phi(0)\big)\right].}}

We can write
\eqn\vand{
\prod_{j<k}^{J-M} (z_j-z_k)^2 = \sum_f 
\left| \matrix {
1       & z_{f_1}   & \cdots & z_{f_1}^{J-M-1} \cr
z_{f_2} & z_{f_2}^2 & \cdots & z_{f_2}^{J-M-1} \cr
\vdots  & \vdots    & \ddots & \vdots          \cr
z_{f_{J-M}}^{J-M-1} & z_{f_{J-M}}^{J-M} & \cdots & z_{f_{J-M}}^{2J-2M-2} \cr
} \right| }
and Taylor expand $X(z_i)$ around $z_i=0$
\eqn\taylex{
e^{-i\sqrt 2 X(z_i)} = e^{-i\sqrt 2 X(0)}
\left[ \sum_{k=0}^{\infty} S_k \left( \{ {-i \sqrt 2 \over k!} \del^k X(0)
\} \right) z_i^k \right]. }
In \vand\ the sum is over all permutations $f=(f_1,...,f_{J-M})$ of
$(1,2...,J-M)$. Putting  \vand\ and \taylex\ into \psieva,
and using the symmetry of the integrand over the index $i$, we have

\eqn\allpsi{
\psi_{J,M}^\pm  \sim
\left|\matrix {
S_{2J-1} & S_{2J-2}   & \cdots & S_{J+M} \cr
S_{2J-2} & S_{2J-3}   & \cdots & S_{J+M-1} \cr
\vdots   & \vdots    & \ddots & \vdots \cr
S_{J+M}  & S_{J+M-1}  & \cdots & S_{2M+1} \cr
} \right|
Exp \left[\sqrt 2 \big(iMX(0) +
(-1 \pm J) \phi(0)\big)\right] }
with $S_k = S_k \left( \{ {-i \sqrt 2 \over k!} \del^k X(0) \} \right)$ and
$S_k = 0$ if $k<0$.
We will denote the rank $(J-M)$
determinant in \allpsi\ as $\Delta(J,M,-i \sqrt 2 X)$.
As a by-product, comparing the two definitions in \disc\ we can use
\allpsi\ to deduce a mathematical identity relating
the determinants of rank $(J-M)$ and $(J+M)$,
\eqn\mathe{\Delta(J,M,-i \sqrt 2 X) = (-1)^{J+M+1} \Delta(J,-M,i \sqrt 2 X).}

We now begin to study the DGS. One first notes that the DGS in
\pgstwo\ can be generated by
$\int {dz\over 2 \pi i} e^{-\sqrt 2 \phi(z)} \psi^-_{\half, \pm \half}(0)$.
In general
it is also possible to write down explicitly one
of the many gauge states for each
discrete momentum in the $\psi^-$ sector as follows
\eqn\gsmm{\eqalign{
G^{-}_{J,M} &\sim
\left[\int {dz \over 2 \pi i} e^{-\sqrt 2 \phi}(z)
\right] \psi^-_{J-1,M} \cr
&\sim S_{2J-1}(\{ {-\sqrt2 \over k!} \del^k \phi \})
\Delta(J-1,M,-i \sqrt 2 X) e^{iMX + (-1-J)\phi}.}}
Using \seten\
it can be verified explicitly after a length algebra
that they are primary, and are of dimension 1. For $M=J-1$
\gsmm\ are ``pure $\phi$'' states, but orthogonal to the ``pure X''
discrete physical states at the same momenta, and are therefore
gauge states. For general $M$ the polynomial prefactor in \gsmm\
factorizes into ``pure $\phi$'' and ``pure $X$'' parts,
and are still orthogonal to
the physical states at the same momenta. They are, therefore,
also gauge states.
This is also suggested by the following result \refs{\kp}
\eqn\winftm{
\int {dz\over 2 \pi i}
\psi_{J_1,M_1}^- \psi_{J_2,M_2}^- \sim 0 }
where the r.h.s. is meant to be a DGS.
We thus have explicitly obtained a DGS for each $\psi^-$ discrete momentum.
We stress that there are still other DGS in this sector, for example,
the states
\eqn\gspphi{
G'^-_{J,M} \sim \left[\int {dz \over 2 \pi i}
e^{- \sqrt 2 \phi (z)}\right]^{J-M} \psi^-_{M,M}}
can be shown to be of dimension 1. Since they are ``pure $\phi$''
states, they are also DGS.
This expression reminds us of \disc . However, there is no $SU(2)$
structure in the $\phi$ direction, and the usual techniques of $c=1$ 2d
conformal field theory cannot be applied.
The ``pure $\phi$ DGS'' are only found in
the minus sector.

For the plus sector, the operator products of
the discrete states defined in \disc\ form a $w_\infty$ algebra \refs{\kp},
\eqn\winftp{ \int {dz\over 2 \pi i}
\psi_{J_1,M_1}^+ \psi_{J_2,M_2}^+ = (J_2 M_1 - J_1 M_2)
\psi_{J_1+J_2-1,M_1+M_2}^+  .}
(Again, the r.h.s. is up to a DGS.)
We can subtract two positive norm discrete states to obtain a pure
gauge state as following
\eqn\gspps{\eqalign{
G^{+}_{J,M} = & (J+M+1)^{-1}
\int {dz \over 2 \pi i} \left[\psi^+_{1,-1}(z) \psi^+_{J,M+1}(0) +
\psi^+_{J,M+1}(z) \psi^+_{1,-1}(0) \right]  \cr
\sim & (J-M)!
\Delta(J,M,-i\sqrt 2 X) Exp\left[\sqrt 2\left(iMX+(J-1)\phi\right)\right] \cr
& +(-1)^{2J} \sum_{j=1}^{J-M} (J-M-1)! \int {dz \over 2 \pi i}
{\cal D}(J,M,-i\sqrt 2 X(z),j) \cr
&Exp\left[\sqrt 2\left(i(M+1)X(z)+(J-1)\phi(z)-X(0)\right)\right]
}}
where ${\cal D}(J,M,-i\sqrt 2 X(z),j)$ is defined as
\eqn\cald{{\cal D}(J,M,-i\sqrt 2 X(z),j) =
 \left|\matrix {
S_{2J-1} & S_{2J-2}   & \cdots  & \cdots& S_{J+M} \cr
S_{2J-2} & S_{2J-3}   & \cdots  & \cdots& S_{J+M-1} \cr
\vdots   & \vdots    & \ddots  & & \vdots \cr
(-z)^{j-1-2J} & (-z)^{j-2J} &  & & (-z)^{j-J-M-2} \cr
\vdots   & \vdots    &  & \ddots& \vdots \cr
S_{J+M}  & S_{J+M-1}  & \cdots  & \cdots& S_{2M+1} \cr
} \right| ,}
which is the same as $\Delta(J,M,-i\sqrt 2 X(z))$
except that the $j^{th}$ row is replaced by
$\{ (-z)^{j-1-2J},(-z)^{j-2J}... \}$.
As an example, with \gspps\ one easily obtains the state
$G^+_{{3\over 2}, \pm \half}$ of \dgtwopl .

\newsec{$w_{\infty}$ Charges and Conclusion}

It was shown \refs{\kp} that the operators products of the states
$\psi^+_{J,M}$ defined
in \disc\ satify the $w_{\infty}$ algebra in \winftp .
By construction \gspps\
one can easily see that the plus sector DGS $G^{+}_{J,M}$ carry
the $w_{\infty}$ charges and can be considered as the symmetry
parameters of the theory.
In fact, the operator products of the DGS $G^+_{J,M}$ of \gspps\
form the same $w_{\infty}$ algebra
\eqn\gswin{\int {dz\over 2 \pi i} G^+_{J_1,M_1}(z) G^+_{J_2,M_2}(0)
=(J_2 M_1 - J_1 M_2) G^+_{J_1+J_2-1,M_1+M_2}(0)}
where the r.h.s. is defined up to another DGS.

In summary, we have shown that the space-time $w_{\infty}$ symmetry
parameters of 2D string theory come from solution
of equations \gauone\ and \gautwo .
This argument is valid also in the case of 26D (or 10D) string theory
although it would be very difficult to exhaust all the solutions of
the gauge states \refs{\lee}. This difficulty is,
of course, related to the high dimensionality of the space-time.
The DGS we introduced in the old covariant quantization in this paper seems
to be related to the ghost sectors and the ground ring strucrture \refs{\wit}
in the BRST quantization of the theory. Many issues remain to be studied.


\vskip 1cm

\noindent
{\bf Acknowledgements.}

The typesetting of this paper has used computers of
the Department of Applied Mathematics of NCTU.
TC would like to thank H. Verlinde and S. Kachru, and
JL would like to thank A. Jevicki, M. Li
and B. Vrosevic for discussions in the early stage of this work.
This research
is financially supported by National Science Council of Taiwan,
R.O.C., under grant number NSC84-2811-M009-006 and NSC83-0208-M-009-063.

\vskip 1cm
\vfill\supereject
\listrefs

\end